\newcounter{myctr}

\documentclass{ws-acs}
\usepackage{cite}
\usepackage{graphics}
\begin{document}

\markboth{D. Helbing {\em et al.}}{Ant Behavior Under Crowded Conditions} 

\catchline{}{}{}{}{}
%
%%%%%%%%%%%%%%%%%%%%%%%%%%%%%%%%%%%%%%%%%%%%%%%%%%%%%%%%%%%%%%%%%%%%

\title{ANALYTICAL AND NUMERICAL INVESTIGATION OF ANT BEHAVIOR UNDER CROWDED
CONDITIONS}

\author{\footnotesize Karsten Peters$^1$, Anders Johansson$^1$,
Audrey Dussutour,$^{2}$ and Dirk Helbing$^{1,3}$}
\address{$^1$ Institute for Transport \& Economics, Dresden University of Technology,\\
Andreas-Schubert-Str. 23, 01062 Dresden, Germany\\[1mm]
$^2$ Biology Department, Concordia University,\\ 
7141 Sherbrooke St. W., Montreal, Queubec, H4B 1R6, Canada\\[1mm] 
$^3$ Collegium Budapest~-- Institute for Advanced Study,\\
Szenth\'{a}roms\'{a}g u. 2, 1014 Budapest, Hungary}

\maketitle

\begin{history}
%\received{(received date)}
%\revised{(revised date)}
%\accepted{(Day Month Year)}
%\comby{(xxxxxxxxxx)}
\end{history}

\begin{abstract}

Swarm intelligence is widely recognized as a powerful paradigm of
self-organized optimization, with numerous examples of successful applications in
distributed artificial intelligence. However, the role of physical interactions
in the organization of traffic flows in ants under crowded conditions has only
been studied very recently. The related results suggest new ways of congestion
control and simple algorithms for optimal resource usage based on local
interactions and, therefore, decentralized control concepts. Here, we present
a mathematical analysis of such a concept for an experiment with two alternative
ways with limited capacities between a food source and the nest of an ant colony. Moreover, we carry
out microscopic computer simulations for generalized setups, in which ants have
more alternatives or the alternative ways are of different lengths. In this way and
by variation of interaction parameters, we can get a better idea, how powerful
congestion control based on local repulsive interactions may be. Finally, we will
discuss potential applications of this design principle to routing in traffic or
data networks and machine usage in supply systems.
\end{abstract} 


\section{Introduction}\label{sec:intro} 

A basic principle in ant trail formation is the indirect interaction and 
communication of ants through the environment by deposited chemical pheromones
({\it stigmergy}) \cite{Gra59,Wil62}. It enables ant colonies to make
adaptive choices based solely on local information. Together with travel
time experiences (e.g. in {\it Linepithema humile} 
\cite{Gos89}) or effects of U-turns (e.g. in {\it Lasius niger} \cite{Bec92})
it leads individual ants to the use of the shortest among
several paths to transport food from distant places into their nest. 
\par
The stigmergy concept in general allows for a complex collective
cooperative behavior of relatively simple agents \cite{Dor00a} and has
inspired a large number of new algorithms and applications especially in
the field of combinatorial optimization problems \cite{Dor04}. This
includes shortest path problems \cite{Dor97,Wed04}, routing in
communication networks \cite{Car98} or material flow scheduling on factory
floors \cite{Par00,Val01,Blu05}.
\par
However, efficient transportation on a system-wide level is not only concerned 
with shortest paths, but also with traffic assignment and coping with
bottleneck situations. An insufficient regulation of traffic flow will
usually lead to a congestion of the apparently preferred path. This is
one of the most challenging problems in road traffic and
routing of data on the internet. Generally speaking, 
the efficient distribution of limited resources by decentralized,
individual decisions is still an open problem in many networked systems. 
\par
As will be shown in the following, the investigation of ant behavior can give some hints how to
deal with such problems. Therefore, Secs.~\ref{Sec2} and \ref{Sec3} will focus on ant traffic and 
its load-dependent organization, including some analytical investigations. Section~\ref{Sec4} 
will present some new numerical results for a microsimulation of ants based on the
social force concept. Finally, Sec.~\ref{Sec5} summarizes our results and discusses possible
areas of application.

\section{Transport and Traffic in Ant Societies}\label{Sec2}

Before we study the optimization of ant traffic under crowded conditions,
we shall briefly review some basic properties of the traffic organization in ants.
\par 
Mass recruitment in ants has a vital reason: The
flow of ants should guarantee an efficient return of food to the nest.
This will favor flow maximizing trail systems and traffic organization. 
However, another objective is the minimization or limitation 
of round trip times $\tau$ (i.e. the time a 
single ant needs to reach a food source from the nest and to return). 
Any delay incurred by an ant during its round-trip slows down 
the amplification mechanism underlying the recruitment process. In situations 
of competition for food sources even a small decrease in round-trip duration could 
improve the fitness of  a colony against a 
competitor by rapidly building up a worker force at the food source. A shorter round trip 
time for the same flow of food return
therefore increases the performance and flexibility and, hence, the
fitness of an ant colony.
\par
Thus, we expect that ant behavior supports an optimal traffic organization
in terms of the flow $\Phi$ and round-trip time $\tau$. It is no
surprise that these performance criteria also apply to routing algorithms
in information networks and that they are meaningful for the transport of goods as well.
\par 
In some regards, ant traffic is comparable to
vehicular traffic or pedestrian behavior \cite{Hel01}. Laden, nestbound
ants experience a slower speed than unladen, outbound ants. Therefore, an
ant trail may be congested with large variations in individual speed and
agility. However, ants cooperatively driven by shared goals of the colony
are different from the rather selfish behavior and distinct individual
interests in human transport systems.
\par
In contrast to vehicular traffic,
which is organized in unidirectional lanes of traffic, a complete directional
separation for outbound and inbound ants is uncommon due to the chemical
(chemotaxis-based) trail attraction. Whereas in some ant species like the wood ant {\it Formica
rufa} \cite{Hol55}, some termites \cite{Miu98}, army ants \cite{Couz03} or 
African driver ants \cite{Got95} forms of unidirectional flow organization
can be observed, in most species a separation of opposite flows does not occur. 
In case of directional traffic, the separation follows certain principles, where
either returning or outbound ants use the middle of a trail and the
other fraction moves at the outer regions of a trail. In 
black garden ants ({\it Lasius niger}) \cite{Bec92} or leaf cutting
ants \cite{Bur03} exploiting renewable or permanent resources, no
separation is observed at all. This may be explained by the large
variations in individual speeds \cite{Hel02}.
\par
Presumably, the absence of directional separation leads to
collisions between ants moving in opposite directions \cite{Bur03}. But in
contrast to larger organisms, ants can stop very abruptly and recover
their initial velocity quickly after a frontal collision with another
ant.
\par
Theoretically, ant traffic flows can be described in the framework
of self-driven many-particle systems \cite{Hel01}. For particle flows in
general and for ants on a trail in particular, we have the formula  
\begin{equation}
\label{eq:flow-density} \Phi=w\rho V \, , 
\end{equation} 
where $\Phi$ is the
flow, $\rho$ the density of ants and $V$ their average speed.
$w$ denotes the width of the trail. However, the relation between $V$ and
$\rho$, known as fundamental diagram, is specific to microscopic
properties of the driven particles and cannot be determined easily from
scratch. Whereas for vehicular traffic and pedestrians a rich body of
theoretical work and empirical measurements exists, not much knowledge has been
gathered on fundamental diagrams of foraging ants \cite{Chowd00,Chowd02, Nis03}.
For {\it Atta cephalotes} (a leaf cutting ant) Burd {\it et al.} found the
empirical speed-density relation \cite{Bur02}: 
\begin{equation}\label{eq:flow-velocity} 
\Phi=w\rho V_m[1-(w\rho/k_m)^n] \, ,
\end{equation}
where $\Phi$ is the flow, $V_m$ represents the maximum average velocity
reached under free flow conditions, and $k_m$, $n$ are empirical
parameters estimated in experiments. For {\it Atta} ants the values of
$V_m=4.04$ cm s$^{-1}$, $k_m=0.59$ cm$^{-2}$ and $n\approx 0.64$ were
confirmed \cite{Bur02,Bur03}. This gives an estimate for the velocity
corresponding to the maximum flow of $V=0.39 V_m$. 

\section{Load-Dependent Traffic Organization in Ants} \label{Sec3}

The basic challenge for any self-organizing traffic optimization is 
the ability to adapt the routes of partial flows to the overall flow situation in the network. 
If one link is congested, a certain portion of the overall flow should use an alternative 
path to reach its destination. The degree of utilization of alternative routes that can be 
called optimal depends crucially on the capacity and weight of the available alternative 
paths and the overall flow that has to be served.  
\par\begin{figure}[htbp]
  \begin{center}
(a)\epsfig{file=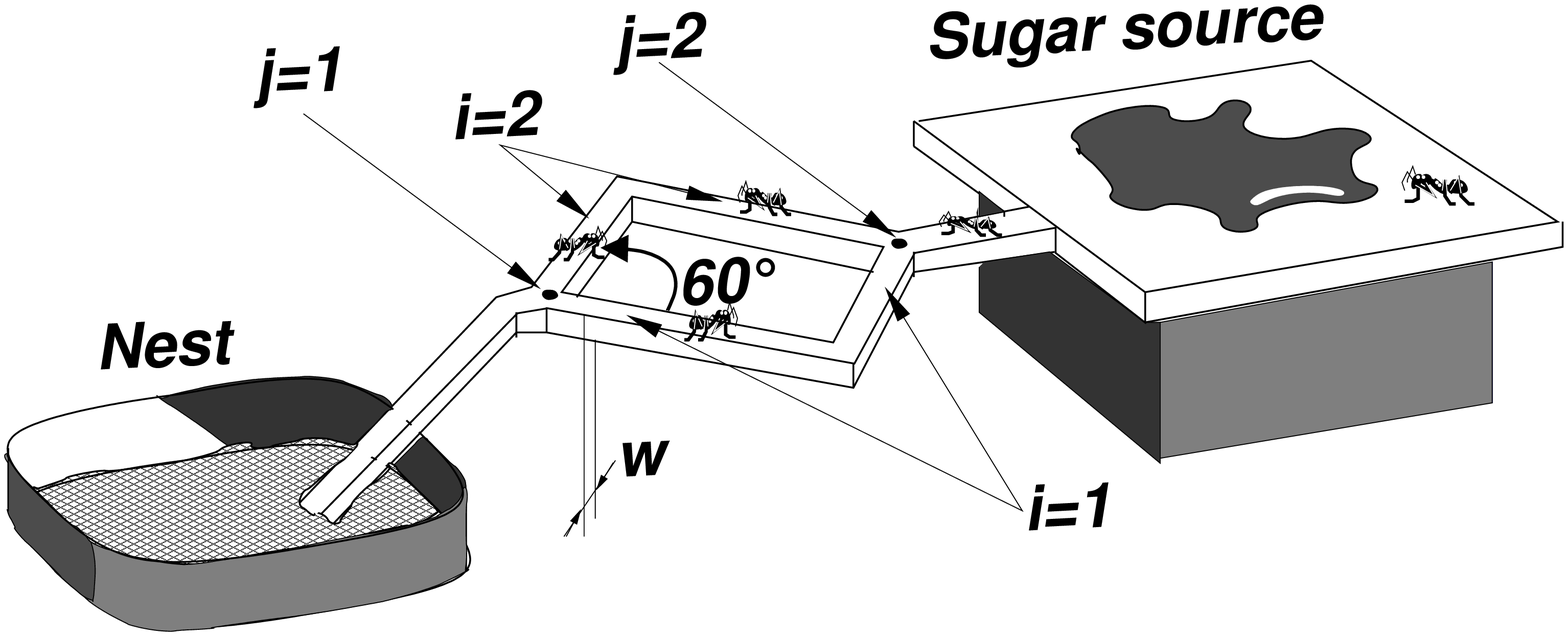, width=8.0cm}\\[5mm]
(b)\epsfig{file=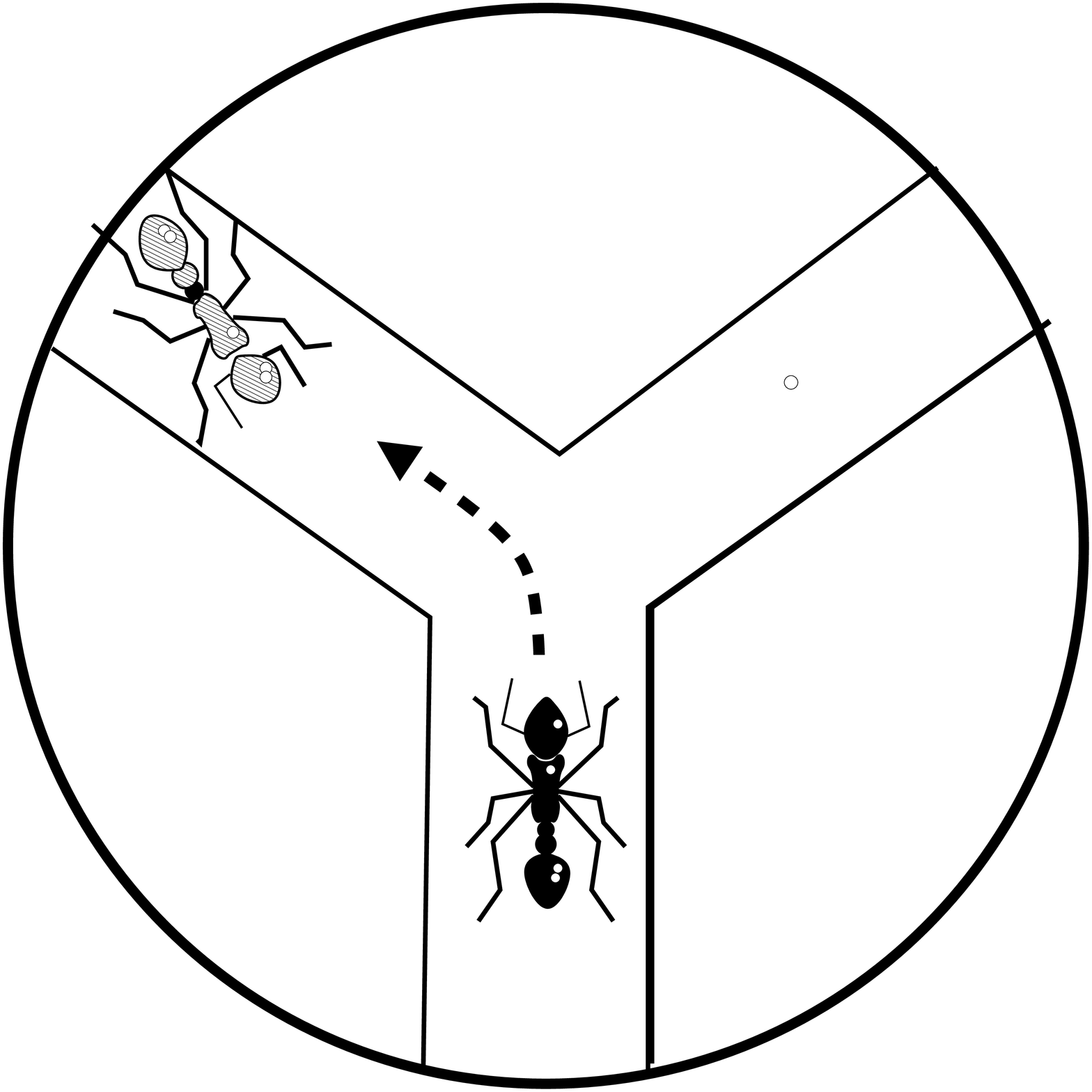, width=2.2cm}
(c)\epsfig{file=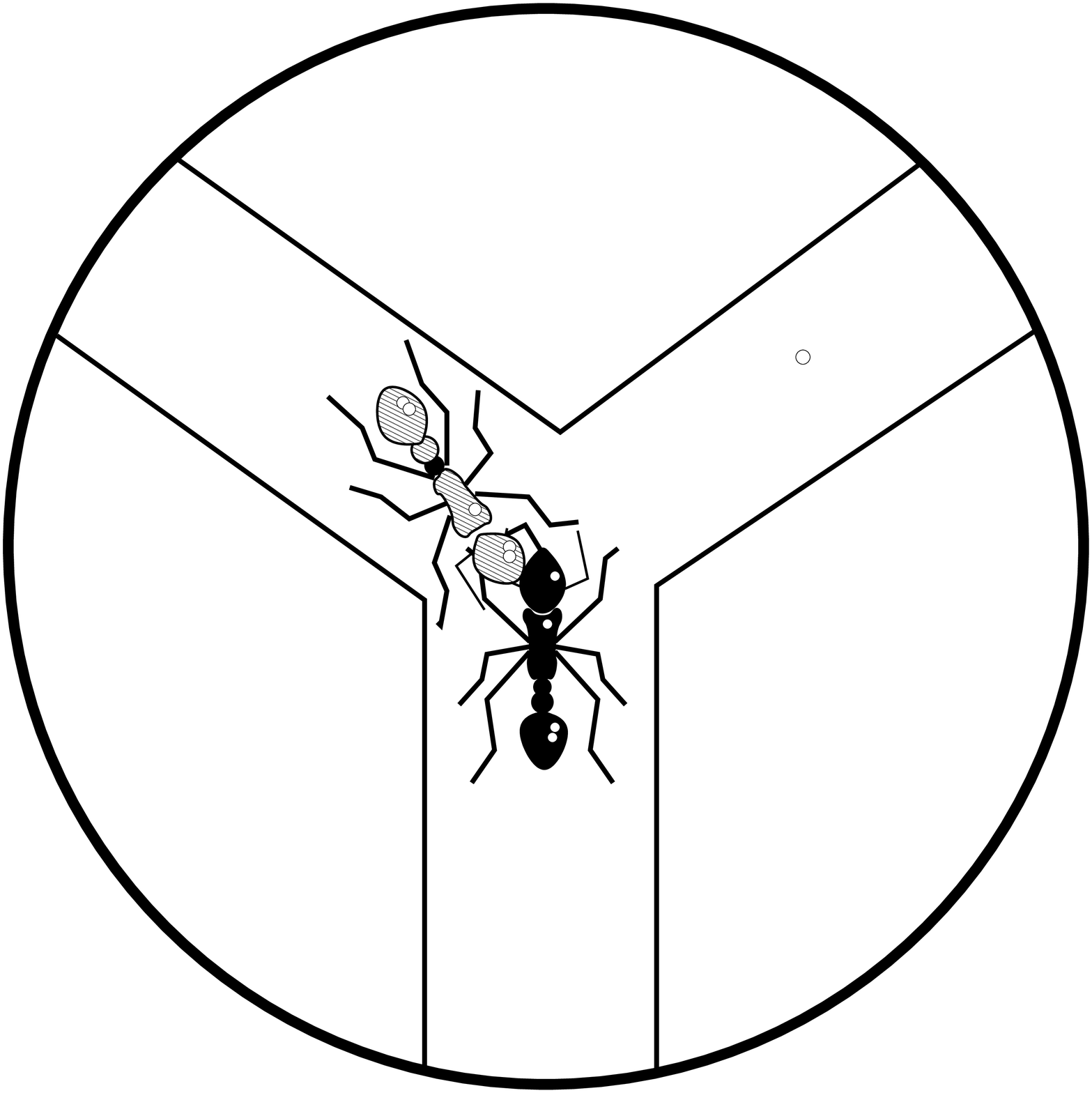, width=2.2cm}
(d)\epsfig{file=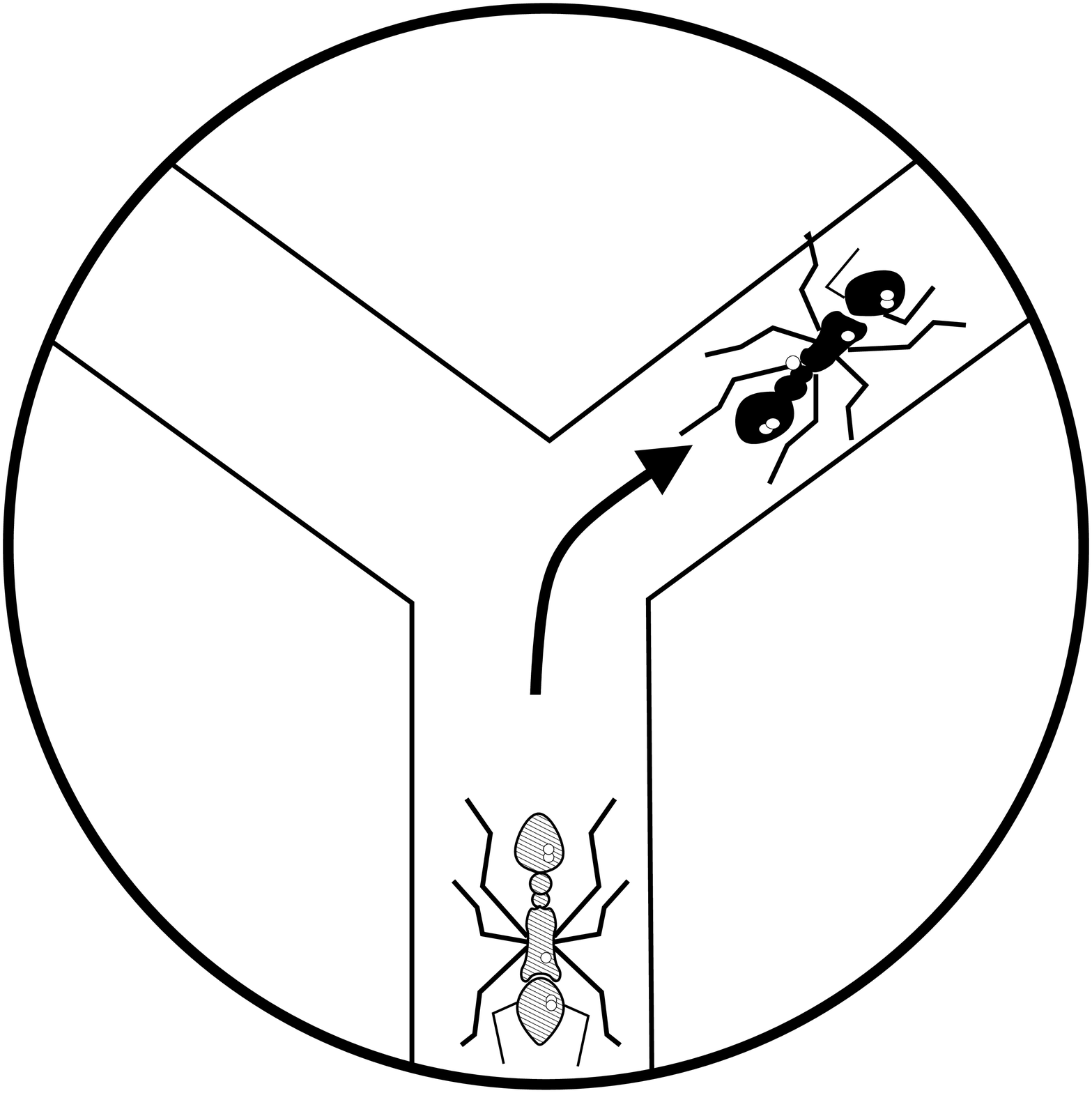, width=2.2cm}
    \caption[]{({a}) Experimental setup of a binary  bridge experiment. 
The setup mimics many natural situations, where the physical constraints 
in the environment  restrict the geometry of trails. Moreover, it resembles 
the situation in technical and economic transport problems, where the geometry 
of networks is usually predefined by the existing infrastructure. To explore different 
capacities of the branches $i=1,2$, the width $w$ of the branches is varied.  
({b}--{d}) Typical encounter at one of the choice points $j=1,2$. 
Two ants, coming from opposite directions collide on a branch near 
the entry point (b). An ant is pushed backwards (c) and redirected towards the 
other branch (d). (After \cite{Dus04}.)}
\label{fig:setup}
  \end{center}
\end{figure} 
To study the organization of ant traffic in a setup with alternative paths
and restricted capacity, the binary bridge experiment by Deneubourg {\em et al.}
\cite{Den90}, in which 
a food source is separated from a nest by a bridge with two alternative, but
equally long branches (Fig.\ref{fig:setup}), can be generalized: According to Eq.
(\ref{eq:flow-density}), the capacity of a branch for ant traffic is
determined by its width $w$. Therefore,  in a recent experiment 
with the black garden ant ({\it Lasius niger}), the widths of both branches
has been varied  from 10mm to 1.5mm to induce increasing levels of
crowding \cite{Dus04}.
\par
It was observed that the recruitment dynamics
was not significantly influenced by the branch width $w$, while the ants
flow organization changed significantly. For wide branches, most ants used
only one of the branches. This spontaneous symmetry breaking can be
explained by small initial fluctuations in the pheromone level. During the
recruitment process, these initial disturbances are amplified, as a
greater number of ants use the trail with the initially or randomly higher pheromone
concentration and reinforce it. Surprisingly, the resulting asymmetric trail usage was not
conserved for widths $w\le 6$mm. Instead, symmetric traffic using both branches
occurred.
\par
This symmetry restoring transition can be explained based 
on head-on encounters of ants traveling in opposite directions \cite{Dus04}. If
the width of the bridges is reduced, the branches become crowded and
the increased number of encounters of ants reduces their average speed. 
Therefore, the restriction to one branch occurs only upto 
a certain critical level of crowding. Above it, head-on encounters at 
the choice points of a narrow bridge  (see
Fig.\ref{fig:setup}b-d) trigger the emergence of a symmetric usage of
both branches: When ants entering one of the branches 
collide with ants from the opposite direction, they are pushed backwards
and, with a probability $\gamma$ of about 60\%, they use the other branch after 
the pushing event at the choice point. While these pushing events were practically
not observed on bridges of 10mm width, they occurred frequently on narrow
bridges \cite{Dus04}. Loosely speaking, a high congestion of ants on one narrow
branch leads to a re-routing of ants through alternative branches.
\par
In the following, we will shortly present a mathematical analysis of this 
result complementary to what has been published in Ref.~\cite{Dus04}. For this,
let $\phi_j =\phi_1$ denote the outbound flow of ants from the nest to the food source and
$\phi_j = \phi_2$ the nestbound flow. Moreover, let $F_{ij}$ denote
the probability to choose branch $i\in\{1,2\}$ at choice point $j\in\{1,2\}$ (see Fig. \ref{fig:setup}a). 
For {\it Lasius niger}, $F_{ij}$ is given by \cite{Bec92} 
\begin{equation} \label{choice}
F_{ij}=\frac{(k+C_{ij})^2}{(k+C_{1j})^2+(k+C_{2j})^2}=1-F_{i'j} \, ,
\end{equation} 
where $F_{i'j}$ describes the probability to choose the
alternative branch $i'=3-i$ at point $j$. The concentration-dependent choice 
begins to be effective if the pheromone
concentration $C_{ij}$ on branch $i$ behind the choice point $j$
exceeds a certain threshold $k$. If every ant leaves a fixed amount $q$ of 
pheromone on its trail and the pheromone decays with a constant rate $\nu$, 
we can determine the local time-dependent pheromone concentration via the equation
\begin{equation}
 \frac{dC_{ij}}{dt} = q [ \Phi_{ij}(t) + \Phi_{ij'}(t-T)] - \nu C_{ij}(t) \, .
\end{equation}
Herein, $T$ is the average time for an ant to get from one choice point to the other
and $\Phi_{ij}$ the overall flow of ants in direction $j$ (behind choice point $j$)
using branch $i$. The formula 
\begin{equation}
 \Phi_{ij}(t) = \phi_j(t) F_{ij}(t) [1- \gamma a \Phi_{ij'}(t-T)/w] 
 + \phi_j(t) F_{i'j}(t) \gamma a \Phi_{i'j'}(t-T)/w 
\end{equation}
takes into account the probability $F_{ij}$ with which branch $i$ is chosen,
but it also considers the effect of pushing events at choice point $j$
to the other branch. $a \Phi_{ij'}(t-T)/w$ is the proportion of ants decelerated
on a branch of width $w$, when the counterflow of ants at
choice point $j'=3-j$ and time $t$ is $\Phi_{ij'}(t)$. The parameter 
$a$ is proportional to the interaction time period and the effective cross section (lateral width) of ants.
$\gamma$ denotes the probability of being pushed to the other branch in cases 
of encounters, i.e. the re-routing probability. One interesting question to be addressed later 
is, whether the empirically found value $\gamma=0.57$ is optimal due to evolutionary
pressure and whether the optimal value depends on the respective scenario.
\par
Due to the normalization $F_{ij}(t) + F_{i'j}(t) = 1$, we have the relation 
$\Phi_{ij}(t) + \Phi_{i'j}(t) = \phi_j(t)$, i.e. $\phi_j(t)$ is the overall flow behind
choice point $j$, as demanded. Furthermore,
in the stationary case we find $dC_{ij}/dt = 0$,
$F_{ij}(t-T) = F_{ij}(t) = F_{ij}$, $\Phi_{ij}(t-T) = \Phi_{ij}(t) = \Phi_{ij}$,
and $\phi_j(t-T) = \phi_j(t) = \phi_{j'}(t) = \phi$ 
(as the nestbound flow and the outbound flow should
be equal). This implies
\begin{equation}
 C_{ij} = \frac{q}{\nu} ( \Phi_{ij} + \Phi_{ij'} ) = C_{ij'} = \dots = \frac{q\phi}{\nu}
 [ \gamma a (\Phi_{i'j}+\Phi_{i'j'})/w + 2F_{ij}(1-\gamma a \phi/w)]
\label{Ce}
\end{equation}
and
\begin{equation}
 F_{ij} = F_{ij'} = 1 - F_{i'j} = 1 - F_{i'j'} \, , 
\end{equation}
because of $C_{ij} = C_{ij'}$ for all $i$ [cf. Eq. (\ref{choice})]. Additionally, with the definitions
\begin{equation}
 \frac{C_{ij}+ C_{i'j}}{2} = \frac{C_{ij'}+ C_{i'j'}}{2} = C \quad \mbox{and} \quad
 \frac{C_{ij}- C_{i'j}}{2} = \frac{C_{ij'}- C_{i'j'}}{2} = D  
\end{equation}
one can show
\begin{equation}
 C_{ij} = C_{ij'} = C + D \, , \qquad C_{i'j} = C_{i'j'} = C - D 
\end{equation}
[see Eq. (\ref{Ce})] and
\begin{equation}
 \Phi_{ij} + \Phi_{ij'} = \frac{\nu}{q} (C+D) \, , \qquad
 \Phi_{i'j} + \Phi_{i'j'} = \frac{\nu}{q} (C - D) \, .
\end{equation}
A detailed calculation gives
\begin{equation}
 C = \frac{C_{ij}+ C_{i'j}}{2} = \frac{q}{2\nu} (\Phi_{ij} + \Phi_{ij'} + \Phi_{i'j} + \Phi_{i'j'}) = \frac{q\phi}{\nu} 
\end{equation}
and, using relation (\ref{Ce}),
\begin{eqnarray}
  D &=& \frac{C_{ij}- C_{i'j}}{2} 
 \nonumber \\
 &=& \frac{q\phi}{2\nu} [ \gamma a (\Phi_{i'j} + \Phi_{i'j'} - \Phi_{ij} - \Phi_{ij'})/w 
 + 2F_{ij}(1-\gamma a\phi /w) -2F_{i'j}(1-\gamma a \phi /w)] \nonumber \\
 &=& - \gamma a \phi D/w + \frac{q\phi}{\nu} (1-\gamma a \phi / w) 
\frac{(k+C_{ij})^2 - (k+C_{i'j})^2}{(k+C_{ij})^2 + (k+C_{i'j})^2} \nonumber \\
&=& C (1-\gamma a \phi / w) \frac{(2k + C_{ij} + C_{i'j})(C_{ij} - C_{i'j})}{(k+C+D)^2 + (k+C-D)^2} 
- \gamma a \phi D / w 
\nonumber \\
&=& \frac{2C(1-\gamma a \phi / w)(k+C)D}{(k+C)^2 + D^2} - \gamma a \phi D / w \, .
\end{eqnarray}
This allows one to calculate all values of $C_{ij}$, $\Phi_{ij}$, and $F_{ij}$ with the above formulas.
We find
\begin{equation}
\label{eq:Csolution}
C_{ij}=C_{ij'}=\frac{q \phi}{\nu}+D \quad \mbox{and} \quad C_{i'j}=C_{i'j'}=\frac{q \phi}{\nu}-D,
\end{equation}
with 
\begin{equation}\label{eq:D=0} 
D=0 
\end{equation} 
or 
\begin{eqnarray}
 D^2 &=& \frac{2C(1-\gamma a \phi / w)(k+C)}{1+\gamma a \phi / w} - (k+C)^2 \nonumber \\
&=& C^2 - \frac{k^2 + \gamma a \phi k^2 / w + 4\gamma a \phi k C / w 
+ 4\gamma a \phi C^2 / w}{1+\gamma a \phi / w} \nonumber \\
&=& \frac{q^2\phi^2}{\nu^2} - \frac{k^2+\gamma a \phi(k+2q\phi/\nu)^2/w}
{1+\gamma a \phi / w} 
\, . \label{eq:Dsquare} 
\end{eqnarray}
As long as $D^2>0$, the stable stationary solution is characterized by
an asymmetric usage of the alternative branches with $C_{ij}=\frac{q \phi}{\nu}\pm
\sqrt{D^2}$ and $C_{i'j}=\frac{q \phi}{\nu}\mp \sqrt{D^2}$.
If $\gamma=0$, $D^2>0$ is
fulfilled for $q\phi/\nu>k$. For $q\phi/\nu<k$, the traffic flow is too low to
generate a certain preference, as the pheromone concentration is not high enough to be 
noticed. Then, the analytical solution collapses to $D=0$, resulting in a symmetric usage of 
both branches, as ants choose between both branches at random.
\par\begin{figure}[htb!]
\begin{center}
  \epsfig{file=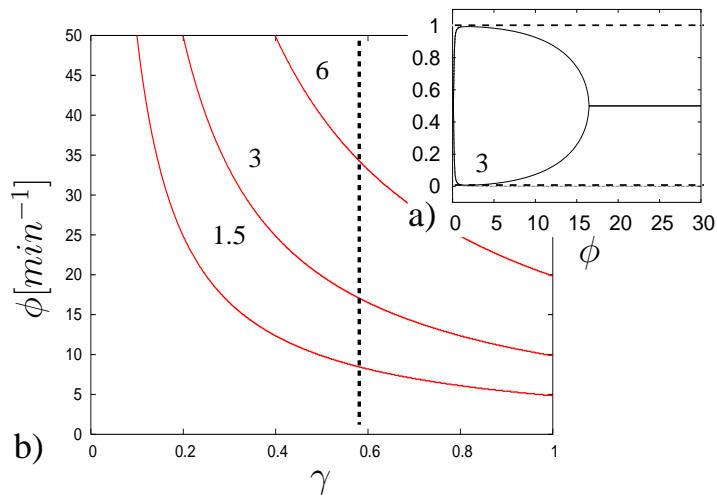, width=12.0cm}
\end{center}     
\caption[]{Analytical results according to Eq. (\ref{eq:Csolution}) for 
the parameter values $q=1$, $k=6$, and $\nu=1/40$min$^{-1}$.
({a}) Fraction of the overall flow on each branch 
in the stationary state, when both branches have a width of 3mm and the value of 
the re-routing probability $\gamma$ is 0.57 as
observed for {\it Lasius niger}. The dashed 
curve shows the solution in the absence of pushing ($\gamma=0$), where the 
model predicts asymmetrical traffic for {\em any} flow above a very low value of the overall 
flow $\phi$. (After \cite{Dus04}.)
({b}) Threshold of the symmetry-restoring transition as a function of the overall
flow $\phi$ of ants and the re-routing probability $\gamma$ for branch widths of
$w=1.5$mm, 3mm, and 6mm. The value 
$\gamma=0.57$ found for {\it Lasius niger} is indicated by the dashed line. }
    \label{fig:bifu}
\end{figure} 
While the asymmetric solution is maintained
also for high flows, if no pushing is considered (Fig.\ref{fig:bifu}a), the
situation changes considerably, when a finite re-routing probability $\gamma>0$
is taken into account: 
Above a certain critical threshold $\phi_{\rm S}(w,\gamma)$ 
of the flow, where
$D^2(\phi_{\rm S})$ becomes zero again, a symmetric stationary
assignment of traffic to both branches is observed, which is caused by pushing events.  
The critical flow $\phi_{\rm S}$, at which this symmetry-restoring transition takes place, is a
function of the width $w$ of the branches and the re-routing probability $\gamma$
(Fig.\ref{fig:bifu}b).

\section{Multi-Agent Simulation of Ant Traffic}\label{Sec4}

As the pushing mechanism is based on random encounters in the area of the
decision points only (the nodes of the trail network), it is not clear at all that these would be representative for
the congestion level on the subsequent part of the respective branch, i.e. the whole
link. Therefore, it
is hard to imagine that a mechanism as simple as pushing, based on local interactions,
would lead to an optimal traffic organization. Nevertheless, it could be shown that
the value $\gamma = 0.57$ of the re-routing probability 
caused the usage of an additional branch shortly before
the capacity of a single branch was reached \cite{Dus04}. Was this just accidental, or
is the value of $\gamma$ carefully chosen, e.g. by evolutionary optimization?
\par
\begin{figure}[htbp]
  \begin{center}
\epsfig{file=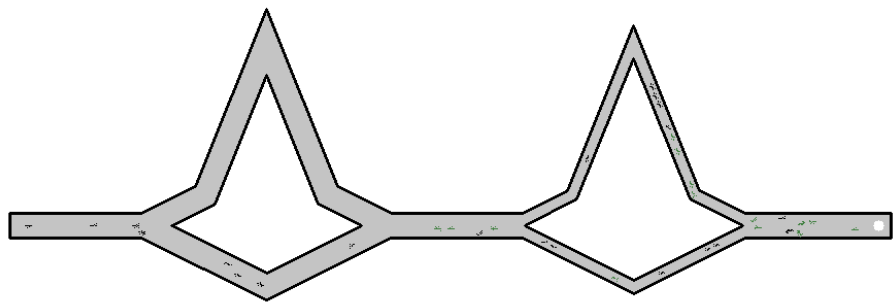, width=13.0cm}
    \caption[]{Simulation setup for a binary  bridge scenario. To explore the effect of different 
capacities of the two alternative branches $i=1,2$, the width $w_i$ of the branches is varied.  
A video film of this scenario clearly shows that
the ants mainly use the shorter of two wide branches, while they use both narrow branches  
(see {\tt http://www.trafficforum.org/ants}).}
    \label{fig:simulation}
  \end{center} \end{figure}

In order to find out whether optimality would also be found for other, 
more complicated  scenarios, we have carried out computer simulations 
(see Fig. \ref{fig:simulation}). Ants were
simulated assuming a distribution of velocities and times at which 
a network with alternative paths (branches)
was entered. Their motion was specified by
a social force model \cite{socforce1,socforce2}, according to which ants had desired directions 
(given by the endpoints of network links) and repulsive interactions. 
The encounters of individual ants were determined, 
and the probability of re-routing due to a
collision near the choice points was given by the parameter $\gamma$.
Alternatively, with a probability of $1-\gamma^{dt}$,
where $dt$ denotes the time discretization of the numerical integration method,
we allowed ants to climb over each other.
\par
Furthermore, in our model we assumed that, once an ant had decided to walk in one
direction on a specific link (branch), it would continue in this direction upto the endpoint of this
link. That is, we neglected U-turns due to interactions. 
When an ant reached a node (choice point), the next destination was chosen among all connected links 
that did not turn more than 90 degrees compared to the ant's previous direction. Stronger turns 
(apart from U-turns) were only considered, if there were no other links available. The choice
probabilities were assumed to be proportional to the squares of the pheromone 
concentrations on the respective links. These were increased by 1 each time an ant passed. However,
before, the pheromone concentrations were multiplied by a certain decay factor in each time step $dt$.
\par
Microsimulations based on this simple multi-agent model (and similar ones) can reproduce the
analytical results in Sec.~\ref{Sec3} quite well (see, e.g., Ref.~\cite{Dus04}). 
Moreover, they allow one to study scenarios in which
\begin{itemize}
\item the branches have different width and/or lengths,
\item there are more than two alternative branches, or
\item parameters such as the re-routing probability are chosen 
different from the empirical value.
\end{itemize}
This facilitates to study the system performance with and without the pushing mechanism
and to relate it to the empirically observed value of $\gamma$.
\par
\begin{figure}[htb!] \begin{center}
\epsfig{file=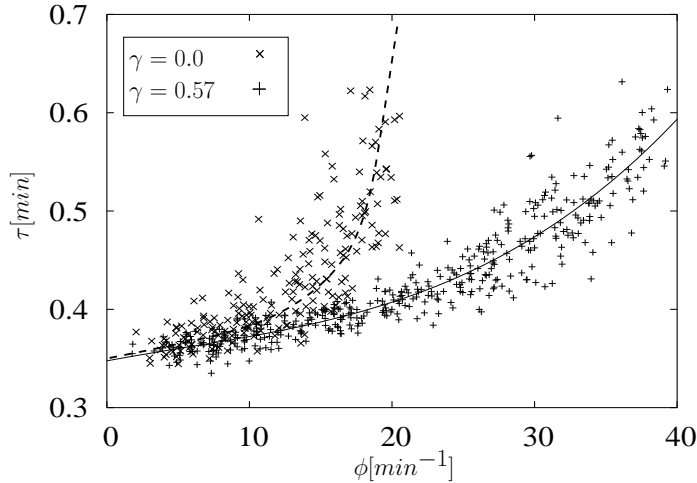, width=10.0cm}
  \end{center}
    \caption[]{Results of Monte Carlo simulations depicting the round trip time vs. 
the overall flow for the binary bridge experiment, taking into account pushing (+) or not (x). 
The curves are fitted as guide for the eyes, while the dashed curve 
represents the case $\gamma=0$, the solid line corresponds to 
simulations considering pushing.\\
A symmetric binary bridge with $w=3$mm of both branches was simulated for 
different flows $\phi$. For each data point (representing the result of one Monte 
Carlo simulation), the values of the round trip time $\tau$ were averaged over 4 minutes after the 
steady state was reached. We used the same parameters as in Fig.\ref{fig:bifu}. The empirically observed 
value of $\gamma=0.57$ was applied to consider the effect of pushing events. Note that
pushing allows to reach higher flow values, as the use of additional branches provides additional
capacity.}
\label{fig:results_max_flow} 
\end{figure}

 
It has been found that the collision-based adaptive routing mechanism can be called
optimal. It seems not only to optimize the flow, but also the round-trip time
by limiting the density-related speed reduction.
Figure~\ref{fig:results_max_flow} shows the averaged round trip time $\tau$ of ants
vs. the reached overall flow $\phi$ as characteristic performance measures
for the quality of a routing algorithm. The binary bridge is assumed to be symmetrical, 
and the parameters applied in the simulation correspond to the findings for
{\it Lasius niger}.  The region where the capacity of one
branch is reached, is characterized by a sharp increase in the round trip
time. Our simulations indicate that, after the symmetry-restoring
transition to the usage of both branches, which occurs
for heavy traffic with $\gamma>0$, the available capacity for returning food 
is doubled compared to a hypothetical
scenario without direct interactions and without pushing ($\gamma = 0$).
\par 
Our simulations also substantiate 
positive effects of pushing ($\gamma>0$) in terms of reduced round trip times. For the same
round-trip time, the interaction-based 
sensitivity to the level of crowding allows a much higher overall flow.
In other words, the observed traffic organization guarantees shorter round
trip times for the same flow of food return compared to hypothetical scenarios without
pushing.
\par
To guarantee the optimality of traffic assignment for a wide
range of flows, the value of $\gamma$ is crucial. It governs the transition
point and balances the influence of the pheromone-driven behavior and
the congestion-related re-organization of flows. For low traffic volumes, 
the beneficial properties of pheromone-guided trail-following behavior
should dominate. Especially for paths of different path lengths, the
majority of ants should be engaged in the shortest path. 
\par
We expect that there is an optimal value of $\gamma$ between 0 and 1: On the one hand,
a high value of $\gamma$ would result in re-routings to longer paths, which is not
necessary if the capacity of the shortest path is not exhausted. However,
in very crowded situations, ants should use additional paths, even if they are longer.
The transition to this behavior should be driven by the utilization of the available capacity
of the preferred branch. On the other hand, if the value of $\gamma$ is too small, 
a transition due to pushing never takes place  (see Fig.~\ref{fig:bifu}b) 
or only, after a undesireable level of congestion has been produced. 
From these considerations we conclude that the optimal value of 
$\gamma$ would trigger the symmetry-restoring transition just before the
capacity of a branch is reached, as has been actually observed \cite{Dus04}. 
\par\begin{figure}[htb!]
\begin{center} 
(a)\epsfig{file=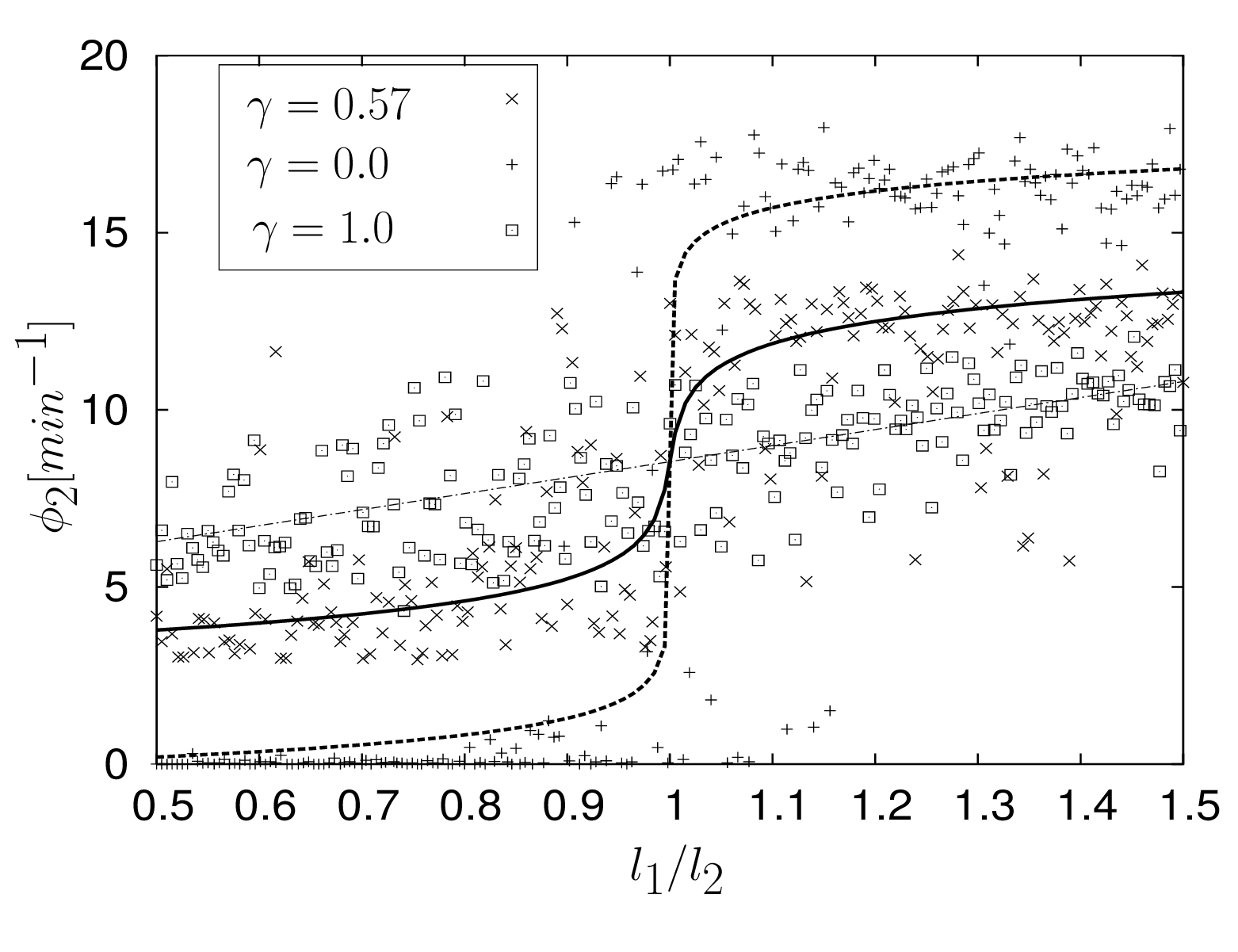,
width=10.0cm}\\ (b)\epsfig{file=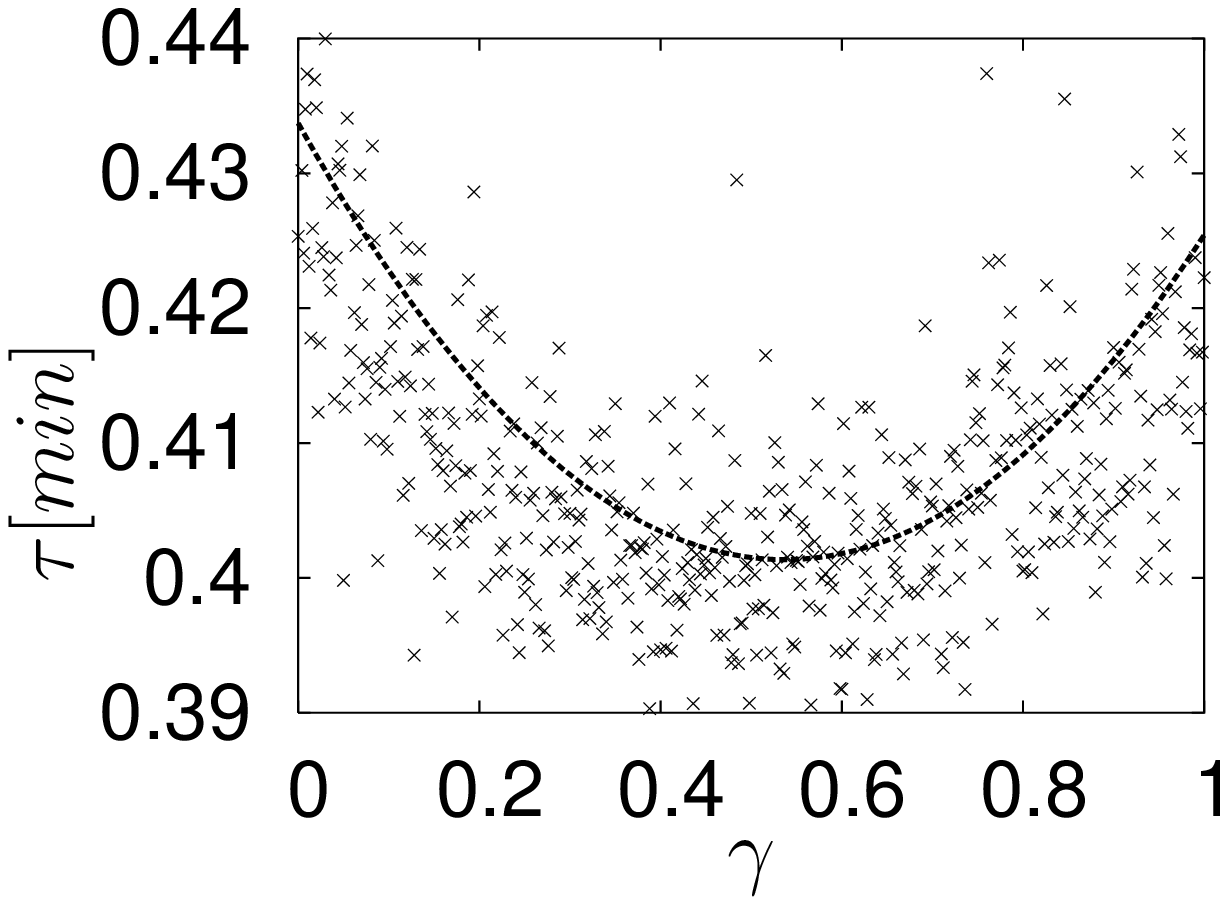,
height=4.2cm}(c)\epsfig{file=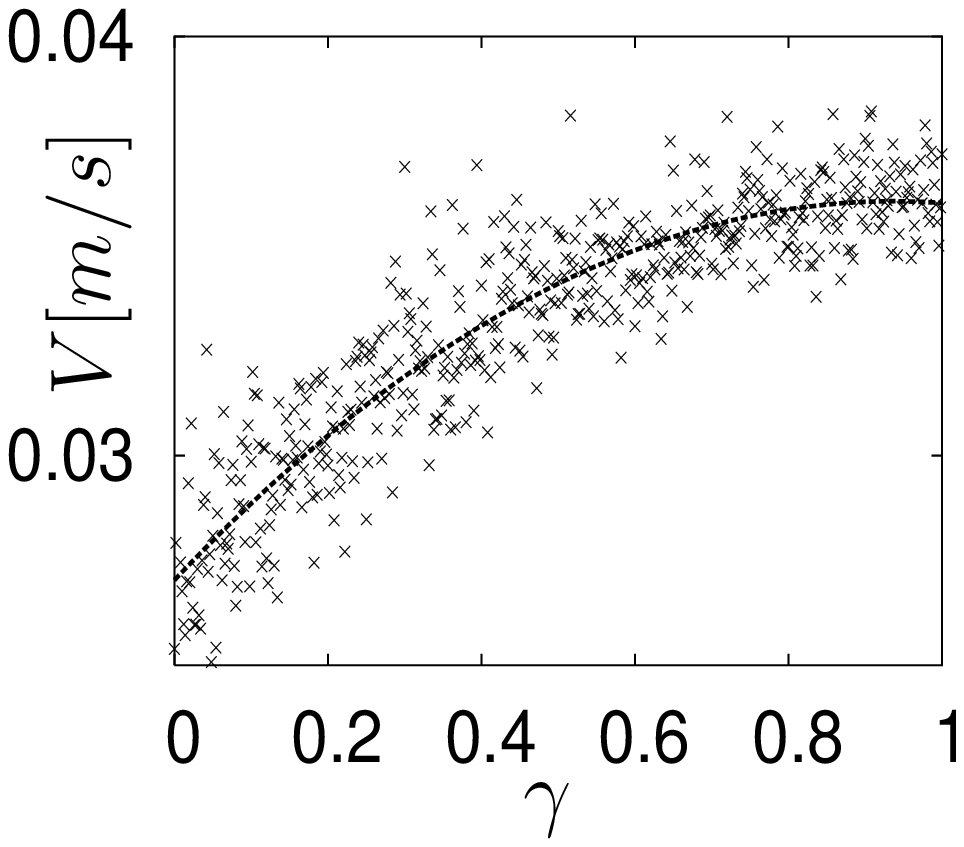, height=4.4cm}
\end{center}
    \caption[]{Results of Monte Carlo simulations for binary bridges with different 
length of the alternative paths. An overall flow $\phi=15\mbox{min}^{-1}$ was applied. 
The length of branch 1 was fixed ($l_1=10$cm), while the length $l_2$ of branch 2
was varied. The simulations were carried out for different values of the re-routing 
probability $\gamma$. Each point represents an average over 10 minutes of simulated ant traffic.
({a}) Fraction $\Phi_{2}$ 
of the overall flow using branch 2 vs. $l_1/l_2$ for 
$\gamma=1$ (every encounter results in a re-routing), $\gamma=0.57$   
(the value observed in {\it Lasius niger}) and without pushing ($\gamma=0$). 
For all values of $\gamma$, the inequality $\Phi_{2}<\Phi_{1}$
for $l_2>\l_1$ reflects the pheromone-driven preference for the shortest path. The curves are 
fitted as guide for the eyes: While the dased curve corresponds to $\gamma=0.0$, the solid 
curve represents $\gamma=0.57$ and the dashed-dotted curve $\gamma=1.0$.
({b}) Averaged round-trip time $\tau$ vs. the re-routing probability $\gamma$ for $\l_1=10$cm, 
$l_2=20$cm and $\phi=\Phi_{11}+\Phi_{21}=15$min$^{-1}$. In this situation, 
the experimentally observed value of $\gamma=0.57$ is the optimal value to 
obtain a minimal round-trip time. A second order polynomial was fitted to the 
simulation results as guide for the eyes.
({c}) Average velocity $V$
vs. $\gamma$. The same parameters as in (b) 
were applied. This plot shows that the re-routing probability $\gamma$ does 
not optimize the velocity, which is maximal if nearly equal flows are assigned 
to both branches.}
    \label{fig:results_gamma_length} 
\end{figure} 
For asymmetric binary bridges with branches of the same width but different
length, the interrelation between the optimal traffic distribution and the
parameters of the local feedback through physical interactions near the choice
points can be discussed also for flows below the capacity of a single
branch (Fig.\ref{fig:results_gamma_length}):
Due to the pheromone-guided behavior, the shorter branch is used more frequently, but with 
increasing $\gamma$, a larger and larger portion of flow utilizes the
longer branch. This effect is observed even if the critical flow for 
symmetry restoring is not reached. For geometrically asymmetric binary
bridges, the symmetry restoring transition is characterized by a 
continuous function 
\begin{equation}
 \Phi_{1j}/\Phi_{2j} \sim l_1/l_2 \, . 
\end{equation}
Thus, the flows beyond
the transition point $\phi_S$ are directly proportional to the lengths of the
alternative paths, but below the transition point, a jump 
in the assignment occurs for the symmetric binary bridge. In Fig.
\ref{fig:results_gamma_length}a, we have $\phi<\phi_S(\gamma)$ for
$\gamma=0$ and $\gamma=0.57$, whereas $\phi>\phi_S(\gamma)$ if
$\gamma=1$.
\par
The optimal assignment depends on the relative length of
both branches and the fundamental diagram of traffic.  In certain
situations we have found numerically that a re-routing probability of
$\gamma=0.57$ provides the optimal behavior in terms of flows and
round-trip times (Fig. \ref{fig:results_gamma_length}b) even for
sub-critical flows ($\phi<\phi_{\rm S}$). However, for other geometries of the
path network, we have found that other values of the re-routing probability
($0.5<\gamma<0.8$) can improve the traffic assignment.
\par
In the same way
as for binary bridges with asymmetric length, traffic organization is
also sensitive to capacity. If two branches $i$ with different widths $w_i$ are
simulated, the flows in the heavy traffic case (after symmetry restoring)
adapt to the capacity of the available branches according to
\begin{equation}
 \Phi_{1j}/\Phi_{2j} \sim w_{2}/w_1 \, . 
\end{equation}
In general, the results obtained in
simulations with different capacities are similar to the traffic organization in
the case of two branches with equal capacities, but different lengths.
\par
\begin{figure}[htb!] 
\begin{center}
\epsfig{file=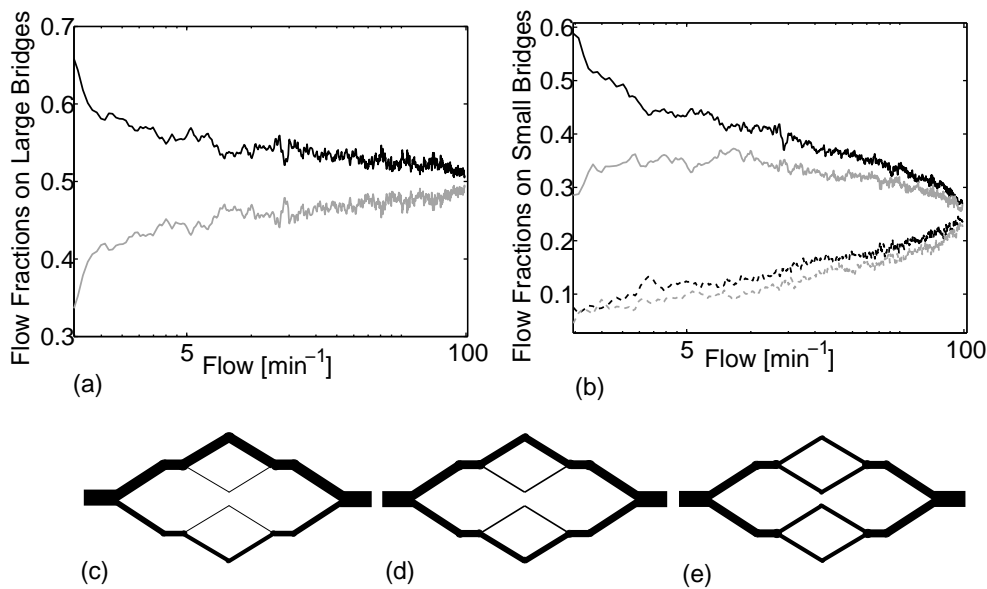, width=13cm}
\end{center}
\caption[]{Simulation results for a completely symmetrical bridge with two large
branches that are again subdivided into two small branches each. The overall width of the
bridge (i.e. all branches) is the same everywhere. (a) Fraction of ants
on the two large branches as a function of the overall ant flow $\phi$. (b) Fraction of
ants on the four small branches. The lower figures give a schematic representation of
the distribution of ant flows over the large and small branches (c) at very low flows,
(d) at medium flows and (e) at large flows.}\label{fourbridge}
\end{figure}
Our microscopic simulations support the conjecture that the discovered
collision-based traffic organization principle in ants generates optimized traffic for a
wide range of conditions. It optimizes the utilization of available
capacities and minimizes round-trip times. Simulation results for a symmetrical bridge with
four branches are shown in Fig.~\ref{fourbridge}. It can be seen that the majority of ants
uses one of the large branches and, within both branches, the majority of ants uses one
of the small branches. Which of the branches is preferred basically depends on random fluctuations
or the initial conditions. When the overall ant flow is increased, the ant flows on the branches
become more equally distributed. Above a certain overall flow, the usage 
of the bridge is completely symmetrical, as for a bridge with two branches. 
On a logarithmic flow scale, 
it appears that we do not have a {\em discontinuous} transition from the use of one small branch
to the use of two small branches and another transition to the use of three or four branches.
However, the sharpness of the transition is certainly a matter of the choice of parameters.

\section{Summary and Outlook}\label{Sec5}

In this article, we have discussed a mechanism of traffic organization in ants
that, on the one hand, establishes the shortest connection based on pheromone attraction,
but on the other hand, guarantees the usage of additional capacities by collision-based
re-routing. In the future, such mechanisms are likely to have interesting applications in the tradition
of ant algorithms and swarm intelligence: For example, if we have a number of
machines with comparable production functions, it will be most profitable to
use one machine only when the production load is small, and to add further machines
when the load exceeds certain thresholds. Note that the optimal thresholds, just as
observed for ants, are usually somewhat below the production capacity
of a machine, as it will in many cases not be advisable to go to machine limits for reasons
of costs, reliability, maintenance, etc. Machine parameters such as load-dependent
cycle times play a role similar to the lengths of alternative paths in
ant systems. 
\par
Altogether, cost-efficient load-balancing of machines or transportation
capacities can be a difficult issue, and interaction-based ant algorithms seem to be a
promising approach, particularly as they do not require central control with the potential
of information losses and delays or decision conflicts. Instead, they are based on simple
local decision rules.
\par
Applications to the routing of traffic are also an interesting subject. Clearly, congestion
can only be avoided as long as there are enough alternative paths over which traffic
can be distributed. If this capacity is not available, the transferability of our ant
model is limited. However, it appears that congestion in data and vehicle traffic could 
be reduced by better routing algorithms. The interaction-based approach by ants would
be promising here as well, although generalizations will be needed due to 
multi-origin-multi-destination flows that are far from being stationary. One
interesting research direction would be the investigation of decision functions different
from Eq.~(\ref{choice}). Another focus could be the extension of the interaction concept, replacing
collisions by passing messages between oppositely moving flow directions. In this way, information
about the traffic volume on different (directed) links could be passed by local interactions
similarly to ad-hoc networks based on transversal hopping, which have been suggested 
for inter-vehicle communication \cite{IVC}. Thereby, it should be possible to establish
a decentralized algorithm for system-wide travel time or congestion information 
(cf. Ref.~\cite{control}), which could be the basis for optimal (re-)routing decisions.

\subsection*{Acknowledgements}

D.H. and A.D. acknowledge the contributions of
Vincent Fourcassie and Jean-Louis Deneubourg to 
the derivation of the analytical results regarding the collisions-based 
symmetry-restoring mechanism proposed in Ref.~\cite{Dus04}. 
K.P. and D.H. thank for partial financial support by the
EU-Project MMCOMNET.

\end{document}